\documentclass[a4paper,11pt]{article}
\usepackage{jinstpub} % for details on the use of the package, please
\usepackage{makecell}
\usepackage{array}
\usepackage[separate-uncertainty,retain-explicit-plus,per-mode=symbol,binary-units]{siunitx}
\DeclareSIUnit\c{c}

\title{\boldmath Muon acceleration at J-PARC}

 \author[a]{Shusei Kamioka,}
 \author[b]{Masato Kimura}
 \author[a]{and Yuga Nakazawa}
 \affiliation[a]{High Energy Accelerator Research Organization, Ibaraki 319-1106, Japan}
 \affiliation[b]{Graduate School of Science, University of Tokyo, 7-3-1 Hongo, Bunkyo-ku, Tokyo 113-0033, Japan}

\emailAdd{kamioka@post.kek.jp}

\abstract{
Muon acceleration is a key technology for providing low-emittance muon beams over a wide energy range.
Various acceleration schemes have been proposed for applications ranging from low-energy $\mu$SR and precision particle-physics measurements to neutrino factories and
muon colliders. 
Experimental demonstrations of muon acceleration, however, have so far been limited. 
At J-PARC, a positive-muon accelerator based on the production and acceleration of ultraslow muons is being developed, and in 2024 the first radio-frequency acceleration of positive
muons was demonstrated. 
In this review, we provide a brief overview of muon-acceleration methods and related experiments, and then review the acceleration method, current status, and future
prospects at J-PARC.
}

\keywords{Muon acceleration; Ultraslow muon}

\proceeding{ICFA Beam Dynamics Newsletter}

\begin{document}
\maketitle
\flushbottom

\section{Introduction}
\label{sec:intro}
A high-intensity and high-energy muon beam from a muon accelerator opens new opportunities for next-generation muon experiments in both elementary particle physics and materials science~\cite{9789811209604_0010,10.1093/ptep/ptz030, Otani:2021mgc}. 
However, the realization of muon-accelerator-based applications remains challenging. Conventional muon beams originate from the decay of pions produced at a target irradiated by an intense proton beam~\cite{miyake_2010}. They therefore occupy a phase-space volume far larger than the acceptance of a conventional radio-frequency (RF) accelerating cavity. 
Moreover, even if such beams could be accelerated, their large emittance would still limit the achievable beam performance in many applications. 
One of the key challenges in realizing muon acceleration is therefore to prepare, prior to acceleration, a beam with sufficiently small phase-space volume while maintaining sufficient intensity.
The principles and experimental status of muon cooling and moderation are reviewed in a companion article in this issue~\cite{ZhangKhaw_thisissue}.
From the accelerator-design point of view, the short muon lifetime requires both beam preparation and subsequent acceleration to be accomplished rapidly, imposing unique constraints on muon-accelerator design~\cite{Accettura2023,10.1093/ptep/ptac067}.
In addition, for high-energy collider applications, muon decay products also become an important accelerator-design issue. 
Electrons and positrons from beam decays can deposit substantial power in collider-ring magnets and generate backgrounds in detector regions, requiring dedicated shielding and magnet designs~\cite{Alexahin:2011_muon_ir,Accettura2023}.

% MuC after ionization cooling
Muon acceleration schemes have been studied worldwide. 
One major approach toward a muon collider is based on the acceleration of both positive and negative muons after ionization cooling~\cite{Bogomilov2020,Bogomilov2024,Zhu:2025_rectilinear_cooling}.
The ionization cooling technique utilizes the loss of both the transverse and longitudinal momentum of the muon beam in an absorber, followed by reacceleration to restore the longitudinal momentum. 
Beyond this stage, the designs of linacs, recirculating linacs, fixed-field alternating-gradient (FFAG) accelerators, and rapid-cycling synchrotrons have been studied~\cite{Accettura2023}.
Related accelerator concepts have also been developed for a neutrino factory, in which an intense GeV-scale muon beam is stored in a racetrack-shaped storage ring to produce intense and well-collimated neutrino beams from muon decay~\cite{PhysRevD.57.6989, PhysRevSTAB.17.121002}.
Muon-collider concepts are reviewed in a companion article in this issue~\cite{SchulteRogers_thisissue}.

Other muon acceleration schemes have also been proposed.
Electrostatic reacceleration is an established technique for low-energy muon spin rotation ($\mu$SR), in which positive muons moderated to eV energies are accelerated to the keV range~\cite{PhysRevLett.72.2793,MORENZONI2000653,PhysRevAccelBeams.27.054501,TRAGER2000662,doi:10.7566/JPSCP.2.010101}.
Such low-energy muon beams have been routinely used at the Paul Scherrer Institute (PSI) for $\mu$SR measurements.
RF linacs have been developed to accelerate low-emittance muon beams
to the MeV and sub-GeV ranges for precision measurements and materials
science applications~\cite{10.1093/ptep/ptz030}.
A compact cyclotron is also under development to accelerate positive muons to 5~MeV for transmission muon microscopy~\cite{yamazaki:cyclotrons2019-tup024}.
In this scheme, 30~keV positive muons are accelerated to 5~MeV in the cyclotron over approximately 1.1~$\mu$s.
The designed cyclotron has an extraction radius of 262~mm and an average magnetic field of 0.4~T.
Automatic cyclotron-resonance acceleration has been proposed as a
compact method for accelerating muons to the MeV range
~\cite{Otani:2021mgc}.
In a proposed design toward a movable muon accelerator, 10-keV muons are accelerated to 20~MeV over a distance of only 29~cm.
The design assumes a uniform magnetic field of 6.7~T and an 850-MHz RF cavity with a peak power of 5~MW.
These schemes generally assume low-emittance muon beams prepared at low energies in order to match the limited acceptance of conventional accelerator systems.
Plasma wakefield acceleration driven by intense laser or electron beams has also been studied for GeV-scale muon acceleration~\cite{wq5g-mtjq,10.1063/5.0189289}.
Overall, these approaches differ primarily in the preparation of the input muon beam and in its assumed intensity and emittance, as well as in the target energy range and intended application.
Table~\ref{tab:acceleration_schemes} summarizes the representative parameters and characteristics of selected muon-acceleration schemes.

\begin{table*}[htbp]
\centering
\footnotesize
\setlength{\tabcolsep}{3pt}
\caption{
Representative comparison of muon-acceleration schemes.
Emittance values are quoted as reported in the corresponding references.
RLA and RCS denote recirculating linear accelerator and rapid-cycling synchrotron, respectively.
}
\label{tab:acceleration_schemes}
\begin{tabular}{
@{}
>{\centering\arraybackslash}p{2.1cm}
>{\centering\arraybackslash}p{2.0cm}
>{\centering\arraybackslash}p{2.3cm}
>{\centering\arraybackslash}p{1.9cm}
>{\centering\arraybackslash}p{1.4cm}
>{\centering\arraybackslash}p{1.7cm}
>{\centering\arraybackslash}p{2.2cm}
@{}
}

\hline
Scheme
& \makecell{Input/ Final\\energy}
& \makecell{Transverse\\emittance}
& Intensity
& Method
& Status
& Application \\
\hline

PSI LEM~\cite{PhysRevAccelBeams.27.054501}
& $\sim15$ eV/30 keV
& --
& $O(10^{3})~\mu^+$/s
& Electrostatic
& In operation
& Low-energy $\mu$SR
\\

\makecell{J-PARC \\muon linac~\cite{10.1093/ptep/ptz030}}
& \makecell{5.6 keV/\\212~MeV}
& \makecell{$1.5~\pi$ mm$\cdot$mrad\\(norm., total)}
& $O(10^{5})~\mu^+$/s
& RF linac
& \makecell{100 keV\\demonstrated}
& \makecell{$g-2$/EDM,\\microscopy}
\\

\makecell{Compact \\cyclotron~\cite{yamazaki:cyclotrons2019-tup024}}
& 30 keV/5 MeV
& \makecell{$1~\pi$ mm$\cdot$mrad\\(1$\sigma$)}
& $10^{4}$/pulse
& Cyclotron
& \makecell{Under\\development}
& \makecell{Muon\\microscopy}
\\

\makecell{Muon \\ collider~\cite{Accettura2023}}
& \makecell{$\sim20$ MeV/\\multi-TeV}
& \makecell{$\sim25~\mu$m\\(norm., rms)}
& $O(10^{12})$/bunch
& \makecell{Linac/\\RLA/RCS}
& Design
& $\mu^+\mu^-$ collider
\\

\hline
\end{tabular}
\end{table*}

However, beyond electrostatic acceleration, direct experimental demonstrations of RF acceleration of muons have so far been limited. 
An RF acceleration experiment using negative muonium ions ($\mathrm{Mu}^{-}$) was previously demonstrated using an RFQ, accelerating the ions to 89~keV~\cite{PhysRevAccelBeams.21.050101}. 
The $\mathrm{Mu}^{-}$ beam provided a source for a proof-of-principle acceleration test, although its transverse emittance was substantially larger than the acceptance of the RFQ.
As a related model experiment, the non-scaling FFAG concept proposed
for rapid muon acceleration was demonstrated using an electron beam
with EMMA~\cite{Machida:2012_emma}.
In a non-scaling FFAG accelerator, fixed magnetic fields are used while the beam orbit changes with momentum, enabling rapid acceleration over a wide momentum range.
In 2024, the first RF acceleration of positive muons was demonstrated at the Japan Proton Accelerator Research Complex (J-PARC)~\cite{PhysRevLett.134.245001}.

% J-PARC
The J-PARC method for positive muon acceleration takes a different route from the collider-oriented schemes~\cite{10.1093/ptep/ptz030, 10.1093/ptep/ptac067}.
This method aims to produce a high-quality positive muon beam with a normalized total transverse emittance of 1.5~$\pi$ mm$\cdot$mrad and an intensity of more than $10^5$~$\mu^+$/s.
This scheme employs ultraslow muons, which have a much smaller normalized emittance than conventional muon beams, and the accelerator is designed to accelerate them to an appropriate energy for each application with limited emittance growth and decay loss.
This technique is well-suited for precision measurements of the muon anomalous magnetic moment ($g-2$) and searches for the muon electric dipole moment (EDM)~\cite{10.1093/ptep/ptz030}. 
In materials science, a transmission muon microscope has also been proposed, offering high-resolution imaging of thick samples~\cite{NagataniAPMC13,TmuM}. 
In addition, $\mu$TRISTAN, a $\mu^+\mu^+$ collider, has been proposed as a future application of this technique~\cite{10.1093/ptep/ptac059}.

In this paper, we review the method, current status, and future prospects of the muon-acceleration program at J-PARC, including the first demonstration of RF acceleration of positive muons.

\section{Muon acceleration method at J-PARC}
\label{chap:method}
\begin{figure*}[hbt]
    \centering
    \includegraphics[width=0.75\linewidth]{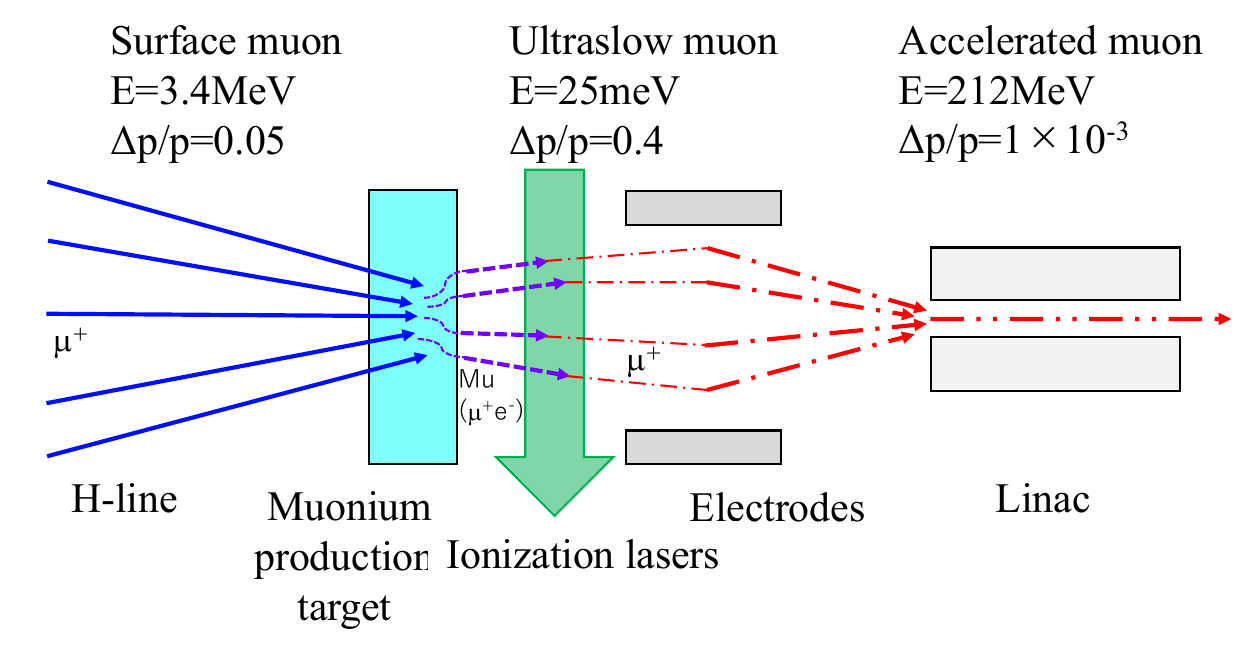}
    \caption{Overview of muon acceleration at J-PARC. Ultraslow muons are produced from a surface-muon beam using an aerogel target and ionization lasers. The ultraslow muons are then electrostatically accelerated and subsequently accelerated by a series of linear accelerators.}
    \label{fig:overview}
\end{figure*}

J-PARC is one of the highest-intensity proton accelerator facilities in the world. A high-intensity positive muon beam with an intensity on the order of $10^8$ $\mu^+$/s is produced at the muon science facility (MUSE)~\cite{miyake_2010} of J-PARC. 
A muon accelerator is being developed using surface muons, produced from the decay of pions at rest, as the primary beam. 
The entire scheme of the muon acceleration is shown in Fig.~\ref{fig:overview}. In this scheme, thermal muons, referred to as ultraslow muons, are produced from a 3.4~MeV surface-muon beam~\cite{PhysRevLett.74.4811}. 
These muons are first accelerated to 5.6~keV with an electrostatic immersion lens.
The resulting keV muon beam is phase-space matched to the acceptance of the downstream linear accelerator and injected into the muon linac system. 

Two important development directions are pursued in this program.
One is to achieve a sufficiently high ultraslow-muon intensity, which is currently determined mainly by the performance of the light source.
The other is the step-by-step installation and commissioning of the accelerator sections while characterizing the beam properties and optimizing the matching between successive sections.

The main elements of this scheme are described below.

\subsection{Production of ultraslow muons}

For the muon-acceleration scheme at J-PARC, a low-emittance muon source is required, since conventional accelerating structures have an acceptance that is orders of magnitude smaller than that of conventional muon beams. For this reason, ultraslow muons are employed as the muon source for the accelerator. Ultraslow muons are thermal-energy positive muons originally proposed in Ref.~\cite{PhysRevLett.74.4811}. As a thermal particle source, ultraslow muons can naturally provide normalized rms transverse emittances below \(1~\pi\mathrm{\,mm\cdot mrad}\).

In this scheme, the surface-muon beam is stopped in a muonium-emission target. 
For a silica-aerogel target, the stopped muon captures an electron to form muonium. 
The muonium then diffuses inside the target and is eventually emitted into vacuum. 
Approximately 1\% of the incident surface muons result in the emission of thermal muonium into vacuum before muon decay~\cite{ZHANG2022167443}.
The muonium emitted into vacuum is ionized using resonant multiphoton ionization. 
For example, pulsed light at 122~nm is used to excite muonium from the ground state to the \(2P\) state, and the excited muonium is subsequently ionized by another light source at 355~nm. Approximately 10\% of the vacuum-emitted muonium is ionized~\cite{doi:10.7566/JPSCP.45.011074}.
Table~\ref{tab:intensity_budget} summarizes the efficiency breakdown of ultraslow-muon production, including the transmission efficiencies in the subsequent acceleration stages. Among these factors, the laser-ionization efficiency depends strongly on the available laser pulse energies, and the value listed here has not yet been achieved experimentally.
Improving this efficiency is one of the main current challenges in achieving a sufficient ultraslow-muon intensity.

The resulting muons are extracted using a static electric field. For transport and phase-space matching of the ultraslow muons to the downstream accelerator, an electrostatic immersion lens, called a Soa lens, is used~\cite{soalens}. 
Approximately 40\% of the ultraslow muons decay during this low-energy transport. 
Because the production time of the ultraslow muons is defined by the short-pulsed laser, the extracted muon beam can have a temporal width as short as about 2~ns.

Starting from the nearly 100\% spin polarization of the surface muon beam, the resulting muon beam is expected to have 50\% spin polarization due to the hyperfine structure of muonium.

\begin{table}[htbp]
\centering
\caption{Approximate intensity budget for the J-PARC ultraslow-muon source and acceleration system.
The laser-ionization efficiency is defined as the fraction of ultraslow muons produced relative to the total number of muonium atoms emitted into vacuum. 
}
\label{tab:intensity_budget}
\begin{tabular}{lll}
\hline
Stage & Approximate efficiency & Basis / Reference \\
\hline
Surface-muon beam
    & -- & $O(10^{8})~\mu^+/\mathrm{s}$,  \\

Muon stopping
    & $40\%$ & Simulation~\cite{ZHANG2022167443} \\

Muonium formation
    & $52\%$ & Simulation~\cite{ZHANG2022167443} \\

Muonium emission into vacuum
    & $6\%$ & Simulation~\cite{ZHANG2022167443} \\
Laser ionization
    & $\sim 10\%$ & Benchmark value~\cite{doi:10.7566/JPSCP.45.011074} \\

\makecell[l]{Low-energy transport,\\including decay loss}
    & $56\%$ & Simulation~\cite{10.1093/ptep/ptz030} \\

\makecell[l]{Transmission through the linac, \\including decay loss} 
    & $72\%$ & Simulation~\cite{10.1093/ptep/ptz030} \\
\hline
Target accelerated-muon intensity
    & -- & $O(10^{5})~\mu^+/\mathrm{s}$ \\
\hline
\end{tabular}
\end{table}

\subsection{Design of the muon linac}
The electrostatically extracted muons are accelerated by a dedicated muon linear accelerator (linac).
A linac-based acceleration scheme is adopted to provide rapid acceleration and suppress muon decay loss within the short muon lifetime of 2.2~$\mu$s.

\begin{figure*}[bt]
    \centering
    \includegraphics[width=0.99\linewidth]{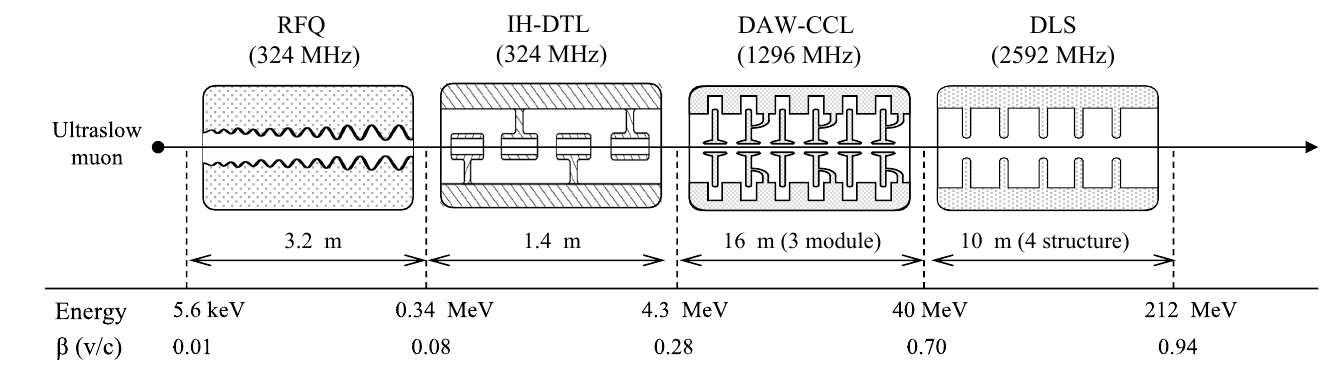}
    \caption{
    Configuration of the J-PARC muon linac and the energy and velocity ranges covered by each accelerating section.
    The RFQ and IH-DTL sections each consist of a single cavity, whereas the DAW-CCL and DLS sections comprise multiple modules and structures, respectively.
    RFQ, IH-DTL, DAW-CCL, and DLS denote radio-frequency quadrupole, interdigital H-mode drift-tube linac, disk-and-washer coupled cavity linac, and disk-loaded structure, respectively.
    }
    \label{fig:muon-linac}
\end{figure*}

The muon linac at J-PARC is designed to accelerate positive muons from 5.6~keV to 212~MeV using four types of accelerating structures optimized for different velocity regions~\cite{10.1093/ptep/ptac067}.
Figure~\ref{fig:muon-linac} shows the configuration of the muon linac.
The accelerator chain consists of a radio-frequency quadrupole (RFQ), an interdigital $H$-mode drift-tube linac (IH-DTL), a disk-and-washer coupled-cavity linac (DAW-CCL), and disk-loaded traveling-wave structures (DLSs).
The entire linac covers a wide velocity range from $\beta = 0.01$ to 0.94 while maintaining efficient acceleration and beam transmission.
Table~\ref{tab:linac_param} summarizes the main parameters of the muon linac.

\begin{table}[bt]
\centering
\caption{Main parameters of the J-PARC muon linac.}
\label{tab:linac_param}
\begin{tabular}{lc}
\hline
Parameter & Value \\
\hline
Injection energy & 5.6~keV \\
Final energy & 212~MeV \\
Velocity range & $\beta = 0.01$--$0.94$ \\
Repetition rate & 25~Hz \\
Normalized total transverse emittance & $1.5~\pi$ mm$\cdot$mrad \\
Momentum spread & 0.1\% \\
Spin polarization & 50\% \\
\hline
\end{tabular}
\end{table}

To preserve the low emittance of the ultraslow-muon beam while suppressing decay loss, comprehensive end-to-end beam-dynamics simulations were carried out for the entire accelerator.
The optimized design is expected to deliver a muon beam with a normalized total transverse emittance of $1.5~\pi$~mm$\cdot$mrad at the linac exit with a momentum spread of 0.1\%.
The beam transmission from the RFQ entrance to the end of the DLS section is estimated to be 98.1\%.
After accounting for muon decay during the total beam transit time of 767~ns, the overall transmission efficiency is expected to be 70.6\%~\cite{phd_ytakeuchi}.

The first accelerating structure is a 324-MHz RFQ linac, which accelerates the muons from 5.6~keV to 0.34~MeV.
The RFQ provides transverse focusing, bunching, and acceleration simultaneously, making it particularly suitable for accelerating ultraslow muons.
A spare RFQ cavity, RFQ II, from the J-PARC H$^-$ linac is reused for muon acceleration~\cite{PhysRevSTAB.16.040102}.
Although the cavity was originally designed for H$^-$ acceleration, it can be adapted for muon acceleration by appropriately scaling the intervane voltage for the muon beam.

Following the RFQ, an IH-DTL accelerates the beam to 4.3~MeV.
The IH structure is suitable for the low-$\beta$ region because of its high effective shunt impedance and compact geometry.
In the muon linac, the IH-DTL adopts the alternating-phase-focusing (APF) method, in which transverse focusing is achieved using a properly designed sequence of synchronous phases without quadrupole magnets inside the cavity.
This approach enables a compact accelerator design without sacrificing beam focusing, making it well-suited for muon acceleration, for which minimizing decay loss is essential.
The full-scale IH-DTL cavity was fabricated, and high-power RF testing was successfully completed, demonstrating stable operation above the design field~\cite{PhysRevAccelBeams.25.110101,ynaka_phd}.

For the medium-$\beta$ region, a 1296-MHz DAW-CCL is employed to accelerate the muons from $\beta = 0.3$ to 0.7, corresponding to energies up to approximately 40~MeV.
The DAW structure combines a high shunt impedance with strong cell-to-cell coupling, enabling efficient acceleration and stable accelerating fields.
Compared with other CCLs, it also has a relatively simple structure and a large coupling constant, providing greater tolerance to fabrication errors.

In the high-$\beta$ region, four 2592-MHz DLSs are employed to accelerate the muons from 40~MeV to the final energy of 212~MeV.
The DLS is a mature technology for electron acceleration and provides a high accelerating gradient.
For muon acceleration, the cell lengths are gradually varied to maintain synchronism between the traveling RF field and the increasing beam velocity. 
The development status of the DAW-CCL and DLS sections is described in Sec.~\ref{sec:beyond-4MeV}.

The muon spin polarization is expected to be preserved throughout the acceleration process~\cite{phd_ytakeuchi}.
Table~\ref{tab:linac_status} summarizes the current status of the J-PARC muon linac development, including the RF acceleration demonstration described in the following section.

\begin{table}[htbp]
\centering
\caption{Current status of the J-PARC muon linac development.}
\label{tab:linac_status}
\begin{tabular}{lll}
\hline
Accelerator section & Energy range & Current status \\
\hline
100 keV RFQ demonstration & 5.7 keV--100 keV &  Acceleration demonstration completed \\
Full-scale RFQ            & 5.6 keV--0.34 MeV & Preparation for installation underway  \\
IH-DTL                    & 0.34--4.3 MeV    &  Preparation for installation underway \\
DAW-CCL                   & 4.3--40 MeV      &  Prototype fabricated; engineering design \\
DLS                       & 40--212 MeV        & Prototype tested; design ongoing\\
\hline
\end{tabular}
\end{table}

\section{Demonstration of RF acceleration of positive muons}
\label{chap:rfq-2024}
The first demonstration of RF acceleration of positive muons was performed in the S2 area, one of the experimental areas at MUSE in 2024. 
Although ultraslow-muon production had long been established, subsequent RF acceleration had not yet been demonstrated. 
The experiment is briefly summarized here; see Ref.~\cite{PhysRevLett.134.245001} for details.

\subsection{Experimental setup}
\begin{figure*}[htbp]
    \centering\includegraphics[width=.8\linewidth]{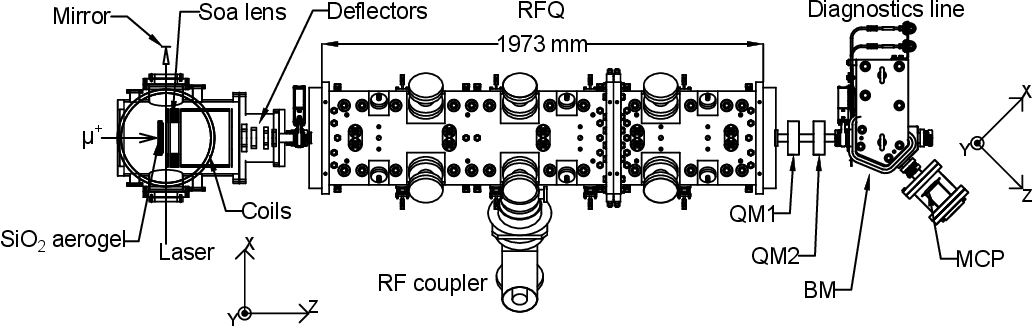}
    \caption{%Schematic drawing of the experimental setup.
    Top view of the experimental setup.
    The surface muon beam is stopped inside a SiO$_2$ aerogel target. The muonium atoms emitted from the target are ionized by a 244~nm laser. The ultraslow muons are transported by the Soa lens and accelerated to 100~keV by a radio-frequency quadrupole (RFQ). Reproduced from~\cite{PhysRevLett.134.245001}. 
    QM, BM, and MCP denote quadrupole magnet, bending magnet, and microchannel plate, respectively.
    }
    \label{fig:setup}
\end{figure*}

Figure~\ref{fig:setup} shows the experimental setup. During this experiment, the surface-muon intensity was $9.5\times 10^{4}$ muons per pulse ($\mu^+$/pulse). 
The surface muons were decelerated by an aluminum foil and injected into a laser-ablated silica-aerogel target~\cite{10.1093/ptep/ptu116, 10.1093/ptep/ptaa145}. 
A pulsed 244 nm light source was used to excite muonium atoms to the 2$S$ state and subsequently ionize them. 
The laser system was originally developed for precision spectroscopy rather than for high-intensity operation, but provided sufficient pulse energy for this demonstration experiment.
The resulting ultraslow muons were accelerated to 5.7~keV and focused onto the entrance of the RFQ by an electrostatic lens system. 
Because of their very low momentum, the ultraslow muons were sensitive to ambient magnetic fields. Helmholtz coils and electrostatic deflectors were therefore used for magnetic-field compensation and trajectory correction.
A prototype RFQ of the J-PARC linac~\cite{hasegawa:linac06-thp072} was used to accelerate the ultraslow muons to 100~keV. 
The properties of the accelerated muon beam were measured with a diagnostics line downstream of the RFQ, where the beam was transported by two quadrupole magnets and analyzed with a horizontal bending magnet. 
At the end of the line, microchannel plate (MCP) detectors were used to measure the particle arrival time and transverse beam profile.

\subsection{Result}

\begin{figure*}[hbt]
    \centering
    \includegraphics[width=0.8\linewidth]{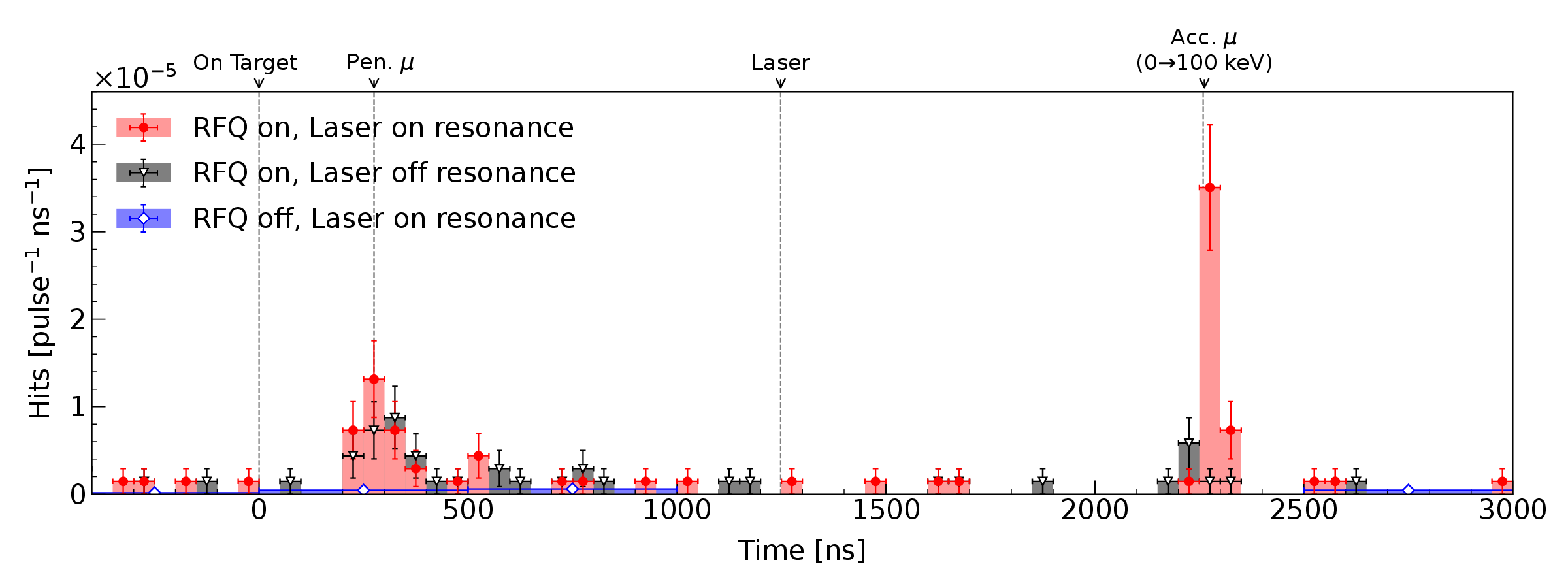}
    \caption{
    Time distributions of the signal pulse under three conditions: without RF power, with RF power on but the laser off resonance, and with RF power and the laser on resonance. The peak at 2280~ns corresponds to the accelerated muon. Reproduced from~\cite{PhysRevLett.134.245001}.
    }
    \label{fig:tof}
\end{figure*}

Figure~\ref{fig:tof} shows the time distributions of the MCP signal where the time origin is the arrival of the surface muon at the muonium production target.
A clear peak was observed at 2280~ns when the RF power was turned on and the laser frequency was tuned to the resonance of the $1S$-$2S$ transition. 
The time of flight agrees with the simulated value of the accelerated muon signal. From these results, it was concluded that the ultraslow muons were successfully accelerated. 
The muon intensity was $ 2 \times10^{-3}$~$\mu^+$/pulse.
The acceleration efficiency, defined as the ratio of the accelerated-muon intensity to the 5.7-keV muon intensity after correcting for muon decay, was estimated to exceed 50\%, indicating that a substantial fraction of the 5.7-keV muon beam was successfully accepted and accelerated by the RFQ~\cite{PhysRevLett.134.245001}.

The measurement of normalized transverse rms emittance was also conducted using the quadrupole scan method. The rms beam sizes after the BM were measured for different quadrupole strengths of a quadrupole magnet. The measured normalized transverse rms emittances of the accelerated muon beam in the horizontal and vertical planes were $(0.85 \pm 0.25 ^{+0.22}_{-0.13} )$~$\pi$ and $(0.32\pm 0.03  ^{+0.05}_{-0.02})$~$\pi~$mm$\cdot$mrad, respectively. 
Compared with the original surface-muon beam, the accelerated beam had a normalized emittance smaller by factors of $2.0\times 10^2$ (horizontal) and $4.1\times 10^2$ (vertical)~\cite{PhysRevLett.134.245001}.
The asymmetry mainly originates from the source size: in the $x$ direction it is $O(10)$ mm, determined primarily by the surface-muon profile, whereas in the $y$ direction it is limited by the 1~mm ionization-laser spot size~\cite{PhysRevLett.134.245001}.
In the full system, the laser spot in the $y$ direction will be expanded, reducing this asymmetry. 
Apart from differences in the surface-muon beamline, the measured $x$-direction emittance is therefore expected to be broadly representative of that in the full linac.

These results provided the first experimental demonstration of RF acceleration of positive muons produced from a low-emittance thermal-muon source, and the measured emittance after acceleration was found to be of the expected order.
The achieved intensity and acceleration energy, however, were still far below those required for practical applications, motivating the developments described in the following section.

\section{Beyond the first RF acceleration demonstration}

\begin{figure}[tb]
    \centering
    \includegraphics[width=0.99\linewidth]{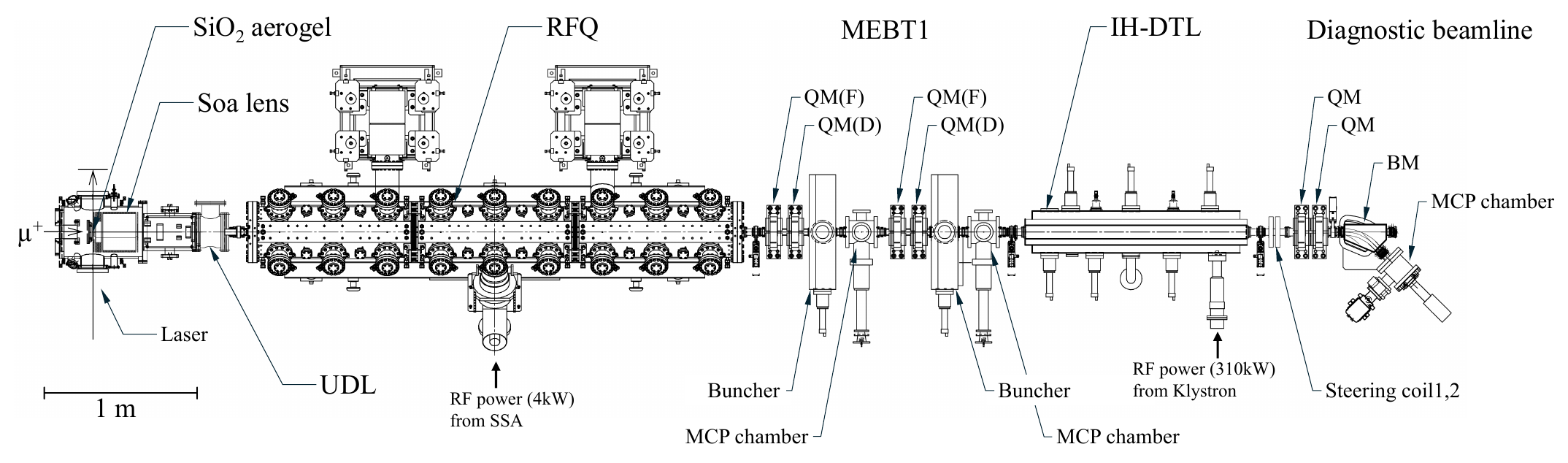}
    \caption{
    Schematic layout of the H2 area for the 4~MeV muon acceleration experiment.
    Surface muons from the H-line are stopped in the SiO$_2$ target, and the emitted muonium is ionized by the lasers.
    The resulting ultraslow muons are transported by the Soa lens and characterized at 5.7~keV in the ultraslow-muon diagnostics line (UDL).
    They are then accelerated by a radio-frequency quadrupole (RFQ), transported through the medium-energy beam transport line (MEBT), and further accelerated to 4.3~MeV by an interdigital H-mode drift-tube linac (IH-DTL).
    QM, BM, MCP, and SSA denote quadrupole magnet, bending magnet, microchannel plate, and solid-state amplifier, respectively.
    }
    \label{fig:h2_layout}
\end{figure}

Following the first RF acceleration demonstration described in the previous section, the next step is to increase the accelerated-muon intensity and extend the acceleration to higher energies.
A new experimental area, the H2 area, dedicated to muon acceleration is currently under development. 
Figure~\ref{fig:h2_layout} shows the layout of the muon linac for the 4~MeV acceleration experiment.
At this stage, multiple accelerator sections and beam-diagnostic systems between them will be installed, enabling beam characterization and matching between successive accelerator sections.

\subsection{Surface muon beamline}

A new muon beamline, the H-line, has been constructed at the Materials and Life Science Experimental Facility (MLF) in J-PARC.
It features a new concept that maximizes the beam intensity up to $\mathcal{O}(10^8)$~\si{\mu^+\per\second}, more than an order of magnitude higher than that used in the S2-area demonstration. 
The H2 area is located on one of the two legs of the H-line.
Details of the design of the H-line can be found elsewhere~\cite{Kawamura:2018apy}.

\begin{figure}[hbt]
    \centering
    \includegraphics[width=0.6\linewidth]{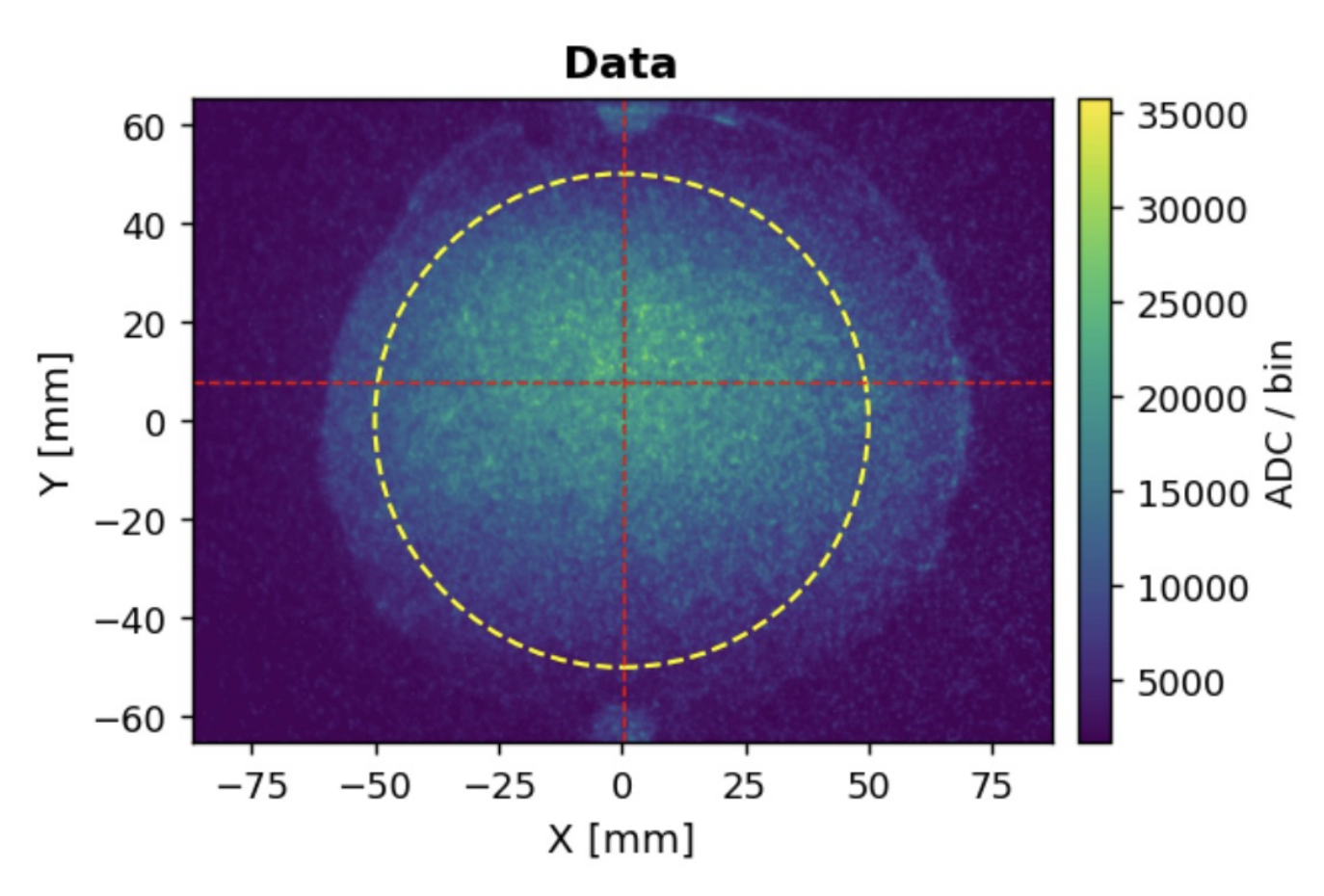}
    \caption{Surface muon beam profile of the H2 area after tuning the beam optics to maximize the ultraslow muon yield.}
    \label{fig:surface_profile}
\end{figure}

The first muon beam was delivered to the H2 area in April 2025.
The beam optics were optimized to form a horizontally elongated surface-muon profile at the aerogel target, improving the overlap with the ionization-laser path. 
The optimized beam profile was measured with a scintillation-plate and CCD-based profile monitor~\cite{Ito:2014jka}, as shown in Fig.~\ref{fig:surface_profile}.
%The beam intensity was measured by stopping the muon beam in an aluminum target and detecting Michel positrons from muon decay, while the polarization was determined using the muon spin-rotation technique.

\subsection{Light source}
\begin{figure}[hbt]
    \centering
    \includegraphics[width=0.55\linewidth]{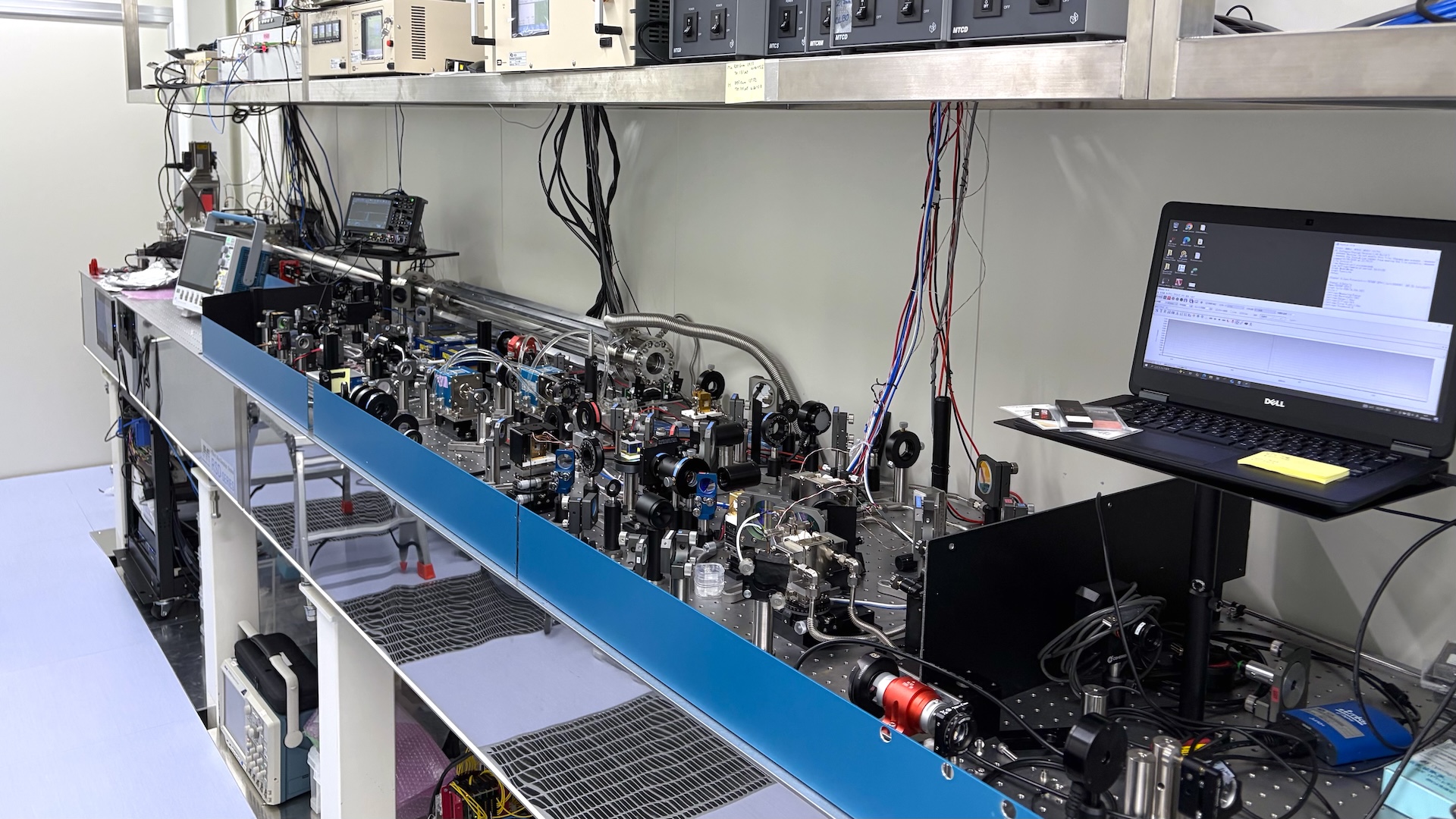}
    \caption{
    A cleanroom adjacent to the experimental area for the development of the muonium ionization light source. 
    The optical table is installed along the cleanroom wall facing the H2 area.
    Laser pulses are transported to the H2 area through an optical port in the radiation shield.
    }
    \label{fig:laser-room}
\end{figure}

To increase the ultraslow-muon yield, the H2 system adopts the 1$S$–2$P$ ionization scheme described in Sec.~\ref{chap:method}, replacing the two-photon 1$S$–2$S$ excitation at 244 nm used in the RF-acceleration demonstration.
Figure~\ref{fig:laser-room} shows the laser room constructed adjacent to the H2 experimental area.

A pulse energy of several $\mu$J at 122~nm has already been demonstrated in a previous study~\cite{Saito:16}.
The H2 system ultimately targets 100~$\mu$J at 122~nm and 300~mJ at 355 nm. 
With these pulse energies, the muonium-ionization efficiency is estimated to reach 10\%, approximately four orders of magnitude higher than the value expected for the RF acceleration demonstration, based on a comparison with simulations under the corresponding laser conditions~\cite{doi:10.7566/JPSCP.45.011074}.

The present laser system has not yet reached the design pulse energies and therefore limits the available ultraslow-muon rate.
The laser system will be progressively upgraded through continued R\&D while acceleration demonstrations are carried out, with the aim of reaching the target pulse energies by the completion of the entire accelerator system.

\subsection{Muonium production and transport of the ultraslow muons}
The same laser-ablated silica-aerogel target used in the RF-acceleration demonstration is employed for muonium production in the H2 area.
The low-energy transport system for ultraslow muons in the H2 area follows the same design concept as that used in the RF-acceleration demonstration. 
The electric field generated by the Soa lens extracts and transports the thermal muons to the entrance of the downstream linear accelerator.
Three orthogonal pairs of Helmholtz coils surround the region and compensate for the ambient magnetic field, while four electrostatic deflectors are inserted between the Soa lens and the linear accelerator for trajectory correction.

\begin{figure}[hbt]
    \centering
    \includegraphics[width=0.4\linewidth]{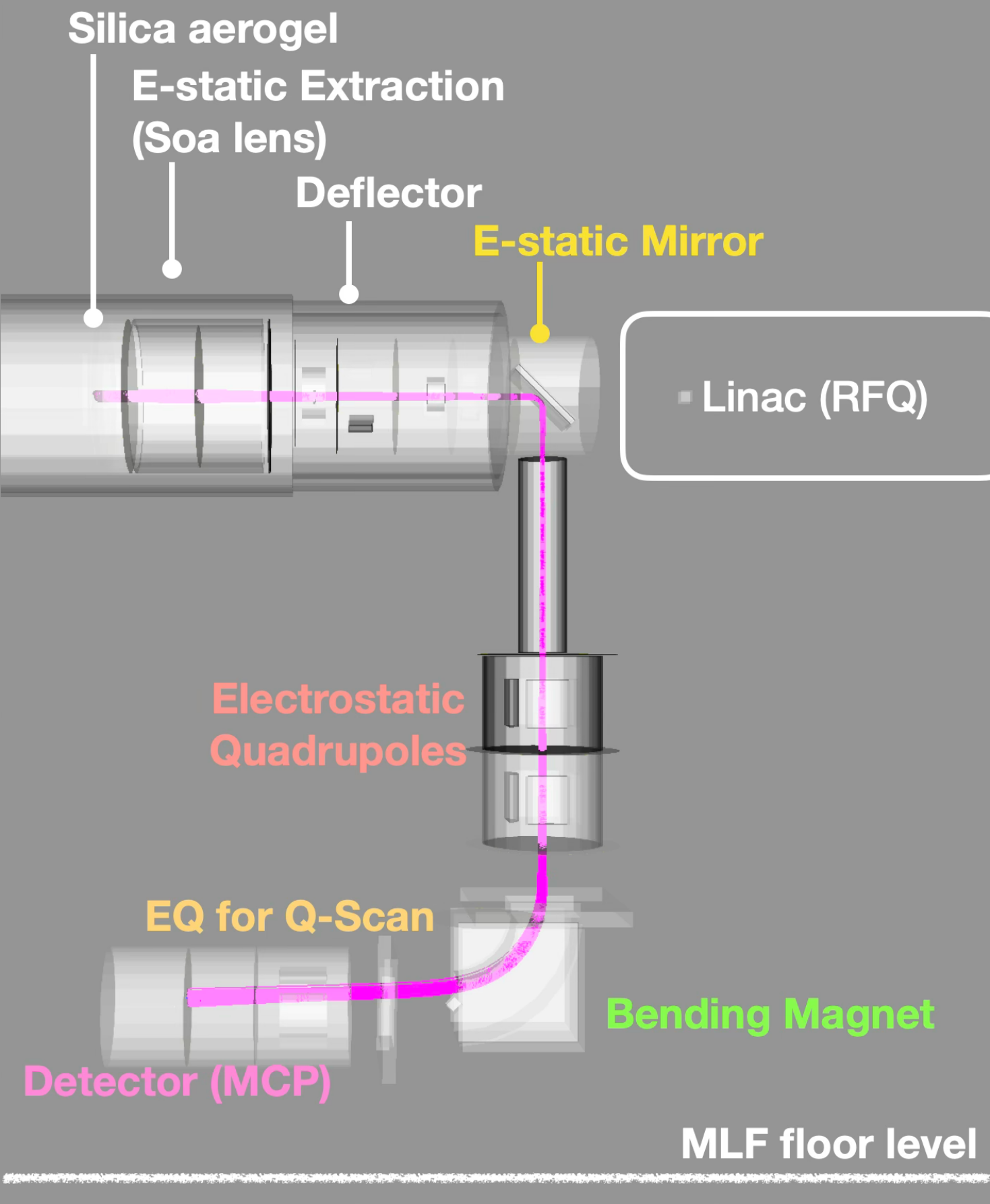}
    \caption{Schematic of the ultraslow muon production apparatus and its diagnostic beamline system.
    The magenta lines represent the simulated trajectory of the ultraslow muon beam.
    EQ, RFQ, and MCP denote electrostatic quadrupole, radio-frequency quadrupole, and microchannel plate, respectively.
    }
    \label{fig:UDL}
\end{figure}

As an additional feature compared with the demonstration setup, the ultraslow muon beam can also be transported to the diagnostic beamline system~(UDL), where the beam intensity and Twiss parameters can be measured~\cite{Wada:2026}.
In particular, measurement of the Twiss parameters is important for beam matching to the accelerator acceptance and for suppressing emittance growth.
The UDL is designed to fit within the limited space just upstream of the accelerator while measuring these parameters with an accuracy of approximately $10\%$.
Figure~\ref{fig:UDL} shows a schematic of the UDL together with the muonium production target.
An electrostatic mirror is employed to deflect the beam toward the diagnostic beamline.
The mirror is sufficiently compact and can be readily retracted from the beamline, allowing flexible switching between the acceleration and diagnostic modes.
An electrostatic-quadrupole~(EQ) doublet and a dipole magnet follow the mirror to transport the beam and select the particle momentum.
An MCP-based beam profiler~\cite{Kim:2018aah} is located at the end of the beamline to measure the beam intensity and profile.

\subsection{4~MeV acceleration}
\begin{figure}[tb]
    \centering
    \includegraphics[width=0.7\linewidth]{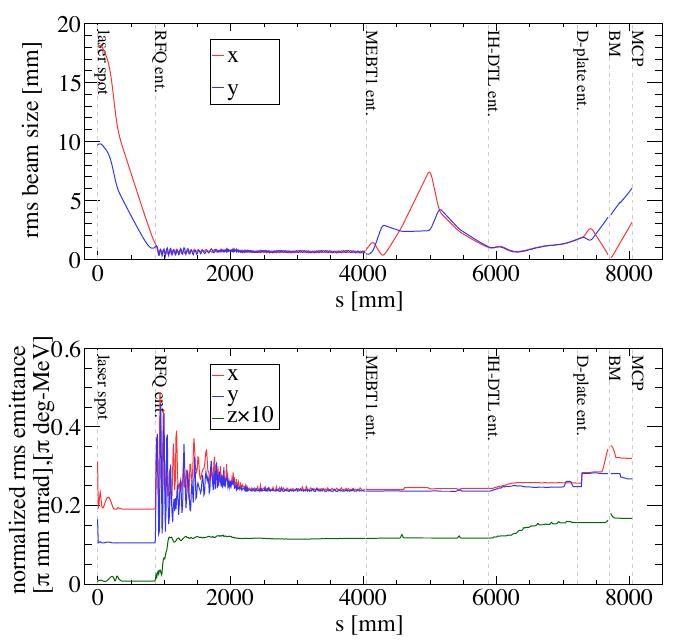}
    \caption{
    %Simulated transverse beam envelope and normalized rms emittance along the beamline.
    %Adapted from Ref.~\cite{Nakazawa_2025}.
    Simulated beam tracking along the beamline.
    The upper panel shows the horizontal ($x$) and vertical ($y$) rms beam sizes as functions of the longitudinal position $s$.
    The lower panel shows the normalized rms emittances in the horizontal ($x$), vertical ($y$), and longitudinal ($z$) directions; the longitudinal emittance is multiplied by a factor of 10 for visibility.
    The vertical dashed lines indicate the positions of the laser spot, radio-frequency quadrupole (RFQ) entrance, medium-energy beam transport line (MEBT1) entrance, interdigital H-mode drift-tube linac (IH-DTL) entrance, D-plate entrance, bending magnet (BM), and microchannel plate (MCP) detector.
    Adapted from Ref.~\cite{Nakazawa_2025}.    
    }
    \label{fig:envelope}
\end{figure}

The major milestone in the H2 area is acceleration to 4.3~MeV using
the full-scale RFQ, a medium-energy beam transport line (MEBT1),
and the IH-DTL.
The accelerated muons will then be transported to a diagnostic beamline, where their bunch time structure and transverse beam profile will be measured.
Figure~\ref{fig:envelope} shows the simulated beam envelope and emittance evolution along the beamline.
The simulations confirm that no significant transverse-emittance growth occurs along the beamline.

%Based on this design, construction and commissioning preparations for the muon linac are in progress.
The RFQ cavity is powered by a 324-MHz solid-state amplifier delivering approximately 4~kW of RF power.
The cavity field is stabilized by a low-level RF feedback system.
MEBT1 will be installed downstream of the RFQ to provide transverse and longitudinal matching to the IH-DTL.
The bunch structure after RFQ acceleration will be measured using a bunch-width monitor based on a microchannel plate detector with high time resolution~\cite{PhysRevAccelBeams.23.022804}.
The transverse beam profile will be measured using the MCP-based beam profiler.
A nominal RF power of 310~kW will be supplied to the IH-DTL by a 324-MHz klystron.
%To reduce construction costs, 
An existing J-PARC klystron will be modified and reused.
A compact Marx-modulator-based high-voltage power supply is also being developed to reduce system size~\cite {takayanagi:ipac2025-wepb062}.
Both the RFQ and IH-DTL will be operated with an RF pulse width of 40~$\mu$s at a repetition rate of 25~Hz, corresponding to a duty factor of 0.1\%.

The available muon-beam intensity will depend on the progress of the light-source development, which is proceeding independently of the acceleration commissioning. 
Although the 4.3~MeV acceleration demonstration itself does not impose a specific beam-intensity requirement, an intensity around \(10^3~\mu^+/\mathrm{s}\) is targeted in the initial stage.

\subsection{Beyond the 4~MeV acceleration}
\label{sec:beyond-4MeV}
\begin{figure}[hbt]
    \centering
    \includegraphics[width=0.7\linewidth]{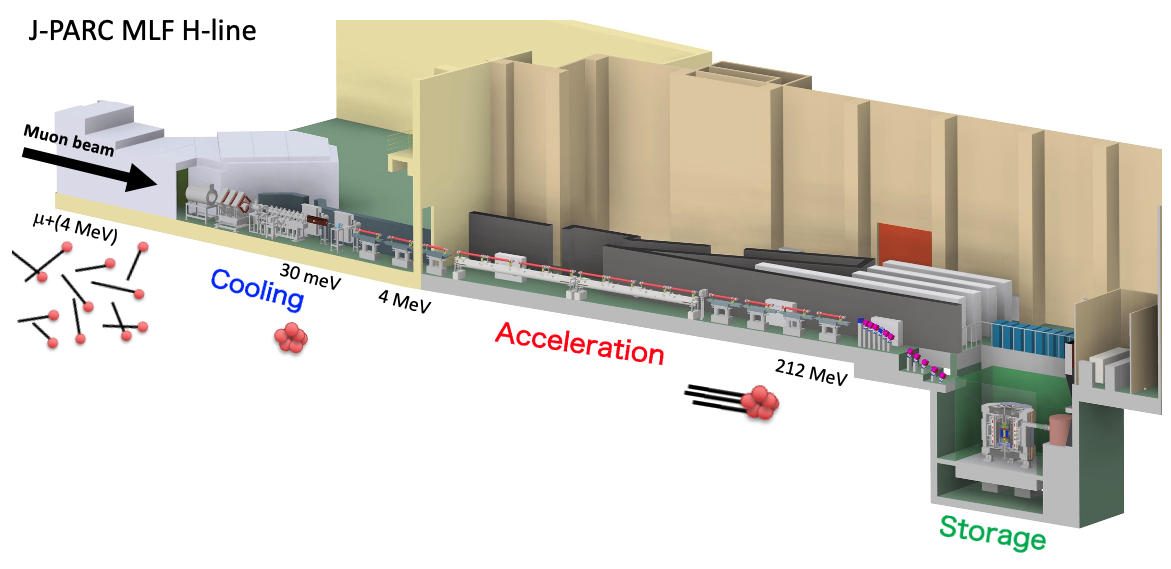}
    \caption{
    Schematic view of the new building planned as an extension of the H2 area for the 212~MeV muon accelerator.
    The 4~MeV accelerator is located in the existing MLF building, while the downstream accelerator sections are planned to be installed in the new building.
    }
    \label{fig:new-building}
\end{figure}

A new building, planned as an extension of the H2 area, is envisioned to house the 212~MeV muon linac.
Figure~\ref{fig:new-building} shows an overview of the new building.
The downstream beamline is planned to branch into two lines, one serving the muon $g-2$/EDM experiment and the other a muon microscope.
A 3~T superconducting storage magnet for the muon $g-2$/EDM experiment is also planned.
This extension will provide the infrastructure for the applications discussed in Sec.~\ref{sec:applications}.
Table~\ref{tab:beam_requirements} summarizes the beam parameters for the 2024 demonstration, the planned H2-area completion, and the full-linac design, together with their associated applications.
The full-linac design specifies a normalized total transverse emittance of 1.5~$\pi$ mm$\cdot$mrad.
Because no corresponding rms-emittance design value is available in the cited references, the rms emittance shown in Table~\ref{tab:beam_requirements} is an approximate value converted from this total-emittance design value and should therefore be regarded only as an approximate value for comparison.

Development of the downstream accelerator sections is in progress.
The DAW-CCL section is designed for acceleration to 40~MeV.
A prototype DAW tank and a coaxial bridge coupler have been fabricated and tested at low power~\cite{phd_ytakeuchi,Kondo_2025}.
The electromagnetic and beam-dynamics designs of the DLS section for acceleration to 212~MeV have been completed, and prototype fabrication is underway~\cite{sumi:ipac2023-mopl172,oai:nagoya.repo.nii.ac.jp:02012455}.

Because the accelerated-muon beam intensity is expected to be of order $10^{5}$~$\mu^+$/s, much lower than typical beam intensities in conventional accelerators, beam monitors with high detection efficiency and the capability to reject accelerator-induced backgrounds, particularly electrons, are required for the downstream sections.
A high-speed bunch-width monitor and a Cherenkov beam-profile monitor using an optical mirror system are under development, with prototype fabrication of the latter underway~\cite{nakagawa:ibic2025-tupco13}.

\begin{table*}[htbp]
\centering
\small
\caption{
Summary of representative beam parameters for the demonstrated and planned stages of the J-PARC muon-acceleration program and their associated applications.
For the rms emittance of the full-linac design, see the discussion in the text.
}
\label{tab:beam_requirements}
\begin{tabular}{lcccccc}
\hline
Stage 
& Application
& Energy
& \makecell{Intensity\\($\mu^+$/s)}
& \makecell{$\varepsilon_{\mathrm{n,rms}}$\\($\pi$ mm$\cdot$mrad)}
& $\Delta p/p$
& \makecell{Spin\\polarization} \\
\hline
2024 demonstration
& -- & 100 keV &  0.05 & 0.85 &  \multicolumn{2}{c}{Not measured} \\

H2-area completion 
&  \makecell{Imaging\\demonstration} & 4.3 MeV & $>10^{3}$ & 0.25 & 0.2\% & 50\% \\

%Muon microscopy
%& muon microscope & 40~MeV &  \multicolumn{4}{c}{Requirements under study}  \\

Full linac
& Muon $g-2$/EDM & 212~MeV& $ O(10^{5})$ & 0.25 & 0.1\% & 50\% \\

\hline
\end{tabular}
\end{table*}

\section{Applications of accelerated positive muon beams}
\label{sec:applications}
Accelerated muon beams with very low emittance enable a range of applications complementary to those targeted by high-intensity collider-oriented beams.

Low-emittance positive-muon beams are particularly attractive for precision measurements that benefit from well-controlled beam phase space, such as measurements of the anomalous magnetic moment ($g-2$) and searches for the electric dipole moment (EDM) of the muon.
Their physics motivations are reviewed elsewhere~\cite{KESHAVARZI2022115675,RevModPhys.91.015001}.
These are storage-ring experiments. 
Conventional muon storage-ring experiments have been performed using muon beams originating from pion decay~\cite{7clf-sm2v}. 
At J-PARC, the use of a low-emittance 212~MeV muon beam enables an independent measurement of the muon $g-2$ with substantially different systematic uncertainties~\cite{10.1093/ptep/ptz030}.
In the proposed scheme, the accelerated muon beam is injected into a compact 3~T storage magnet using a three-dimensional injection method~\cite{Iinuma:2016zfu}.
The experiment aims at a precision of 460~ppb for the muon $g-2$ measurement and a sensitivity of $10^{-20} \mathrm{\,e\cdot cm}$ for the muon EDM.

More recently, a higher-energy option for the muon linac has also been discussed. Since the statistical sensitivity to the muon $g-2$ approximately scales with the Lorentz factor $\gamma$, improved statistical precision can generally be expected with a higher-energy muon beam~\cite{Venanzoni:2025rsd}.

Transmission muon microscopy using accelerated muon beams is another possible application~\cite{NagataniAPMC13,TmuM}. 
A key advantage of muons for transmission microscopy is their large penetration depth in matter, which enables transmission imaging of thick samples.
The low-emittance accelerated muon beam at J-PARC can be focused to a small spot on the sample. 
A typical target sample thickness is 10--100~$\mu$m for a 40~MeV muon beam.
One of the main target applications is the imaging of thick biological samples, such as neuronal cells and neuronal networks.
A compact muon cyclotron is also under development at J-PARC for the acceleration of ultraslow positive
muons to 5~MeV toward transmission muon microscopy~\cite{yamazaki:cyclotrons2019-tup024}.
Within the NCCR Muoniverse program, reacceleration of low-emittance positive-muon beams to the MeV range is also being considered for applications including muon microscopy~\cite{Muoniverse}.

Muon imaging, or muon tomography, is also expected to benefit from accelerated muon beams.
Muon imaging is a powerful tool for the non-invasive inspection of objects~\cite{Bonomi:2020_muon_applications,Tanaka:2023_muography,Schultz:2004_muon_radiography}.
Compared with conventional methods based on cosmic-ray muons, a high-energy, low-emittance muon beam is expected to enable muon imaging with higher spatial resolution and shorter acquisition times~\cite{Otani:2021_muon_imaging,Shimomura:2025_muon_imaging}.
An accelerated muon beam with an energy above 2~GeV and an intensity of $10^3$--$10^4~\mu/\mathrm{s}$ would enable container inspection with an acquisition time of about 40~s per container and a spatial resolution on the order of centimeters.
Such a multi-GeV muon beam is a longer-term possibility beyond the current J-PARC muon-acceleration program.
However, a muon-imaging demonstration using a 4~MeV accelerated muon beam is planned in the H2 area~\cite{Shimomura:2025_muon_imaging}.

With a substantial increase in beam intensity, the low-emittance muon beam could also be applied to future energy-frontier experiments.
Collider concepts based on positive muons have been proposed as a possible route to probe new physics at the TeV scale~\cite{10.1093/ptep/ptac059}. 
Related $\mu^+X$ collider concepts, with $X=e^-$, $\mu^+$, or $p$, have also been discussed~\cite{Akturk_2024,Akturk:2025ubm,Akturk_2025_EPJP}.
One of the main technical challenges is increasing the muon beam intensity beyond $10^{10}~\mu^+/\mathrm{s}$. 
This would require a dedicated muon production target and associated facility infrastructure~\cite{10.1093/ptep/ptac059,Sakaki:2026uvl}.
Such developments would require substantial changes to the overall system and are therefore regarded as a longer-term possibility beyond the current J-PARC muon-acceleration program. 

\section{Conclusion}
Muon acceleration is an essential technology for realizing high-energy, low-emittance muon beams, but direct experimental demonstrations have so far been limited. 
In this review, we summarized the muon-acceleration program at J-PARC, which is based on the production and acceleration of ultraslow positive muons. 
The first RF acceleration of positive muons was demonstrated at J-PARC, establishing an important milestone toward practical muon acceleration. 
Following this demonstration, a new experimental area dedicated to muon acceleration is currently under development, including a surface-muon beamline, a light source for muonium ionization, a low-energy muon-transport system, and accelerating cavities. 
These developments are directed toward multi-stage acceleration of ultraslow muons, a new measurement of the muon $g-2$ and a search for the muon EDM, and applications such as muon microscopy and muon imaging.

\acknowledgments

This work was supported in part by JSPS Kakenhi Grants No. JP18H05226, No. JP19H05606, No. JP20H05625, No. JP21K13944, No. JP21J01132, No. JP22KJ1594, No. JP22K21350, No. JP22H00141, No. JP24H00023, No. JP24K03211, No. JP23K13131, 
No. JP25K17424, 
No. JP26K17183, No. JP26K21728,
No. JP26K21565, %nakazawa wakate
MEXT Q-LEAP JPMXS0118069021, 
JST-Mirai Program JPMJMI17A1,
and 
JST K program No. JPMJKP24J4. 
The authors would like to thank the J-PARC muon section
staff for their support in the conduct of the experiment at
J-PARC MUSE. 
The muon acceleration demonstration experiment was performed at the Materials and Life Science Experimental Facility of the J-PARC under a user program 2011MS06.

\bibliographystyle{JHEP}
\bibliography{reference}

%\begin{thebibliography}{99}

%\bibitem{a}
%Author, \emph{Title}, \emph{J. Abbrev.} {\bf vol} (year) pg.

%\bibitem{b}
%Author, \emph{Title},
%arxiv:1234.5678.

%\bibitem{c}
%Author, \emph{Title},
%Publisher (year).

%\end{thebibliography}
\end{document}